\documentclass{article}
\usepackage{url}
\usepackage{spconf,amsmath,graphicx,booktabs,multirow}
\usepackage{algorithm}
\usepackage{algorithmicx}
\usepackage{algpseudocode}
\usepackage{amsfonts}
\title{CycleTransGAN-EVC: A CycleGAN-based Emotional Voice Conversion Model with Transformer}
\name{Changzeng Fu\textsuperscript{1,3}\thanks{This work was supported by the Grant-in-Aid for Scientific Research on Innovative Areas JP20H05576 (model training), and by JST, Moonshot R\&D under Grant JPMJMS2011 (model evaluation).}, Chaoran Liu\textsuperscript{2}, Carlos Toshinori Ishi\textsuperscript{2,3}, Hiroshi Ishiguro\textsuperscript{1,2}}
\address{\textsuperscript{1}Graduate School of Engineering Science, Osaka University, Japan\\
\textsuperscript{2}Advanced Telecommunications Research Institute International, Japan\\
\textsuperscript{3}Interactive Robot Research Team, Guardian Robot Project, RIKEN, Japan\\
Email: changzeng.fu@irl.sys.es.osaka-u.ac.jp}

\begin{document}
%\ninept
%
\maketitle
\begin{abstract}
In this study, we explore the transformer's ability to capture intra-relations among frames by augmenting the receptive field of models.
Concretely, we propose a CycleGAN-based model with the transformer and investigate its ability in the emotional voice conversion task. In the training procedure, we adopt curriculum learning to gradually increase the frame length so that the model can see from the short segment till the entire speech.
The proposed method was evaluated on the Japanese emotional speech dataset and compared to several baselines (ACVAE, CycleGAN) with objective and subjective evaluations.
The results show that our proposed model is able to convert emotion with higher strength and quality.
\end{abstract}

\section{Introduction}
% Affective expression has burgeoned at a fast speed for artificial intelligence systems, including motions, speech, and facial expressions \cite{sheldon2001virtual, pelachaud2009modelling, chella2008emotional}. 
% Emotional voice conversion (EVC) is one of the important topics in this research field,
% nevertheless, since speech is a complex signal that contains rich information, there is still much room for improving the performance with deep learning methods.
Emotional voice conversion is a special type of voice conversion (VC), 
which aims to transform an utterance's emotional features into a target one while retaining the semantic information and speaker identity. 
Some earlier research in this field focused on mapping the prosody and spectrogram with partial least square regression \cite{helander2010voice}, Gaussian Mixed Model (GMM) \cite{aihara2012gmm, kawanami2003gmm}, and the sparse representation method \cite{ming2016exemplar, takashima2013exemplar}. 
Recently, some researchers leverage deep learning methods to improve the performance of EVC, such as deep neural network (DNN) \cite{vekkot2020emotional, luo2016emotional}, sequence-to-sequence model (seq2seq) with long-short-term memory network (LSTM) \cite{robinson2019sequence}, convolutional neural network (CNN) \cite{kameoka2020convs2s}, as well as their combinations with the attention mechanism \cite{choi2021sequence}. However, these models require to be trained on parallel data, that is, both the source and target should be from the same speaker and have identical linguistic information but in different emotions. 

To reduce the models' reliance on parallel training data, some novel frameworks are introduced into this field. Gao et al. \cite{gao2018nonparallel} proposed a nonparallel data-driven emotional speech conversion method with an auto-encoder. 
Lately, Ding et al. \cite{ding2019group} adopted vector quantized variational autoencoders (VQ-VAE) with the group latent embedding (GLE) for nonparallel data training. 
Moreover, to better learn the mapping function between non-parallel data distributions, cycle-consistent adversarial network (CycleGAN) \cite{zhou2020transforming,liu2020emotional} and variational autoencoder-generative adversarial network (VAE-GAN) \cite{cao2020nonparallel} were introduced to EVC task.
Furthermore, Moritani et al. \cite{moritani2021stargan} employed starGAN to realize non-parallel spectral envelope transformation. 
These EVC models with CNN-based layers trained on non-parallel data all achieved a not bad performance.

Despite the progress made in non-parallel data training, there remains some room to improve the quality of converted emotional voice.
Because speech is a time series with rich acoustic features, there are some interactive temporal relationships among frames need to be considered. 
Although CNNs are well-known for their ability to handle temporal data. To process speech data, which is a sort of lengthy temporal sequence, CNN-based models must be stacked very deep in order to widen the receptive field.
However, the temporal intra-relations would be diluted layer by layer with this manner and make the model suffer from some instability problems such as mispronunciations and skipped phonemes.

\begin{figure*}[t]
\centering
\includegraphics[width=.95\linewidth]{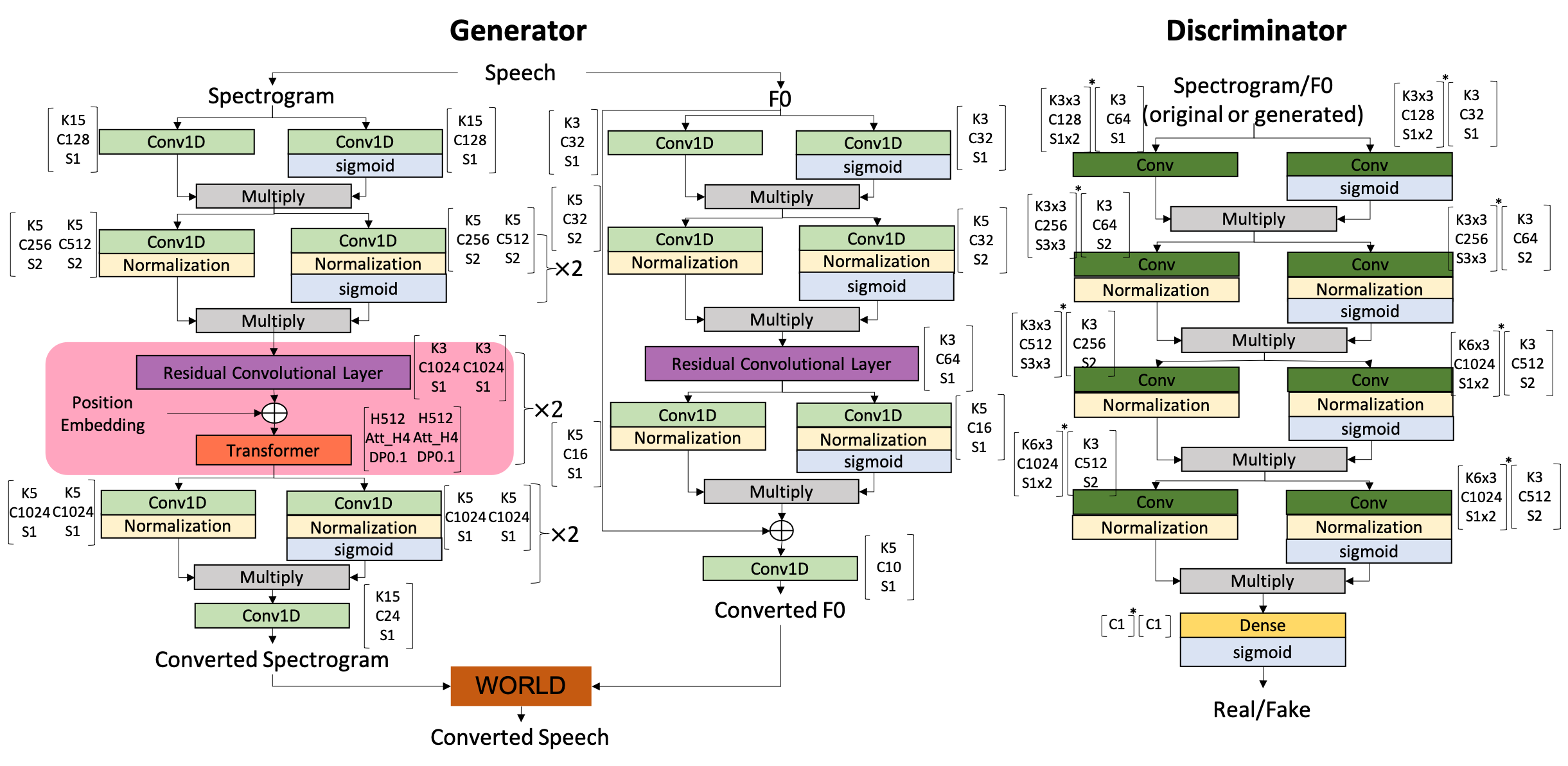}
\caption{Neural networks for the proposed model. $\times n$ means the indicated block repeats n times. The type of $Conv$ in discriminator was variant for spectrogram and F0, $Conv2D$ was used for spectrogram whereas $Conv1D$ for F0. 
The data between $[ \ ]$ are hyper-parameters we set in the experiment. 
$K$ indicates the kernel size, $C$ stands for the number of filters, $S$ is the stride, $H$ represents the number of nods for the hidden layer, $Att \rule{1mm}{0.05mm} H$ is an abbreviation for attention head, $DP$ means the drop out process.$[ \ ]^*$ indicates hyper-parameters for $2D$ processing, $[ \ ]$ for $1D$ processing.}
\label{fig:nn}
\end{figure*}

Considering to augment the receptive field of models and the ability to capture intra-relations among frames, the transformer has been widely discussed in the field of computer vision \cite{dosovitskiy2020image} and natural language processing \cite{wu2020transformer}, and its attention distance is also explored. However, few studies have investigated the capabilities of transformers for the task of speech generation or conversion.
Therefore, to solve the aforementioned problem, in this study:
\begin{itemize}
  \item We proposed a CycleGAN-based model with the transformer and investigated its ability in the EVC task, we named our model CycleTransGAN. 
  \item Moreover, to enhance the model's ability for converting emotional voice, we adopted curriculum learning to gradually increase the frame length during the training.
  \item The proposed method was evaluated on the Japanese emotional speech dataset and compared to several baselines (i.e. ACVAE \cite{kameoka2019acvae}, CycleGAN \cite{zhou2020transforming}) and the different configurations of our proposed model.
\end{itemize}

% This paper is structured as follows. We describe the details of our proposed method in Section 2. The experiment and results are described in Section 3. Discussions are presented in Section 4, and Section 5 briefly summarizes our work.

\section{Proposed Method}

\subsection{Preprocessing}
In this study, we extracted F0 and spectrogram from the speech as our model's inputs inspired by Zhou et al. \cite{zhou2020transforming}, Ming et al. \cite{ming2015fundamental}, and Kaneko et al. \cite{kaneko2018cyclegan}. The following contents demonstrate how the extraction was conducted.

For extracting the F0 feature, the F0 contour was extracted firstly, then, the continuous wavelet transform (CWT) was adopted to decompose it into multiple temporal scales (see Eq.\ref{cwtf0}). $F0(x)$ indicates the input signal, and $\phi$ denotes the Mexican hat mother wavelet. 
\begin{equation}
\centering
W(F0)(\tau, t) = \tau^{-1/2} \int F0(x)\phi(\frac{x-t}{\tau})dx
\label{cwtf0}
\end{equation}
In this study, we set the CWT analysis at 10 scales with one octave apart, which can be represented as:
\begin{equation}
\centering
W_i(F0)(t) = W_i(F0)(2^{i+1}\tau_0, t)(i+2.5)^{-5/2}
\label{cwtf0_1}
\end{equation}
where $i \in [1,10]$ and $\tau_0=5ms$. Finally, the signal is approximately reconstructed as:
\begin{equation}
\centering
F0(t) = \sum_{i=1}^{10} W_{i}(F0)(t)(i+2.5)^{-5/2}
\label{cwtf0_2}
\end{equation}

For the spectrogram extraction, a three-step method so-called CheapTrick \cite{morise2015cheaptrick} was adopted. First, we calculated the power spectrogram based on the windowed waveform as Eq.\ref{CheapTrick1}, the $y(t)$ stands for the waveform, while w(t) indicates the window function:
\begin{equation}
\centering
\int_{0}^{3\tau_0}(y(t)w(t))^2dt = 1.125\int_{0}^{\tau_0}y^{2}(t)dt
\label{CheapTrick1}
\end{equation}
Second, CheapTrick technique smooths the spectrogram with a window of width $2w_0/3$, where $w_0 = 2\pi/\tau_0$:
\begin{equation}
\centering
P(w) = \frac{3}{2w_0}\int_{-w_0/3}^{w_0/3}P(w+\sigma)d\sigma
\label{CheapTrick2}
\end{equation}
After that, the liftering is carried out in the quefrency domain to eliminate the fluctuation:
\begin{equation}
\begin{aligned}
\centering
& P(w) = exp(\mathcal{F}[l_s(\tau)l_q(\tau)p_s(\tau)])\\
& l_s(\tau) = \frac{sin(\pi F0 \tau)}{\pi F0 \tau} \\
& l_q(\tau) = q_0 +2q_1 cos(2 \pi \tau/\tau_0)\\
& p_s(\tau) = \mathcal{F}^{-1}[log(P(w))]\\
\end{aligned}
\label{CheapTrick3}
\end{equation}
where $\mathcal{F}$ and $\mathcal{F}^{-1}$ stands for the Fourier transform and its inversion respectively. $l_s(\tau)$ indicates the liftering function for smoothing the signal, $l_q(\tau)$ stands for the liftering function for spectral recovery. $p_s(\tau)$ represents the Cepstrum of $P(w)$. $q_0$ and $q_1$ were set to 1.18 and -0.09 respectively.

\subsection{Model and Training Strategy}

Given the extracted F0 and spectrogram, we constructed a CycleGAN-based model with the transformer to learn the converting function on non-parallel data (see Fig.\ref{fig:nn}).
% Note that the training procedures for converting F0 and spectrogram were separately conducted.

For converting spectrogram, we first employed the 1-dimension CNN to encode the features. Meanwhile, another CNN branch with sigmoid activation function was designed, and multiply it with the encoded features for selecting the salient ones. 
Then, we inserted a normalization layer after 1-dimension CNN and repeated this CNN-based block two times.
After that, a residual convolutional followed with a transformer layer was designed to capture the temporal relationships among timesteps. The position embedding was added before feeding features to the transformer. This block was also repeated two times.
Subsequently, the CNN-based block with a normalization layer was used to do the feature selection again. 
Finally, we did a post processing with a 1-dimension CNN.

For converting F0, the structure of the neural network was similar to the one for the spectrogram. By considering the quantity of information carried by F0 is much less than that of the spectrogram, we removed the transformer layer and only utilized each block once to reduce the number of trainable parameters.

The CNN-based block was also used in the discriminator. Firstly, we encoded features and selected the salient ones. Then, inserted a normalization layer after the CNN layer and reused this block three times. Finally, a dense layer with a sigmoid activation was employed to output the real/fake label. 
Note that the types of CNN utilized in the discriminator were different for spectrogram and F0; for spectrogram, the 2-dimensional CNN was used, while the 1-dimensional CNN was employed for F0. Furthermore, the proposed discriminator not only produced a label at the utterance level, but also gave multiple outputs (fine-grained level) that presented the real or false samples according to the frames to determine how close each frame was to the real samples.

\begin{table}[t]
\centering
\caption{Information of dataset.}
% For LaTeX tables use
\begin{tabular}{lllll}
\toprule
 & Neutral & Happy &  Anger & Sad  \\
\noalign{\smallskip}\hline\noalign{\smallskip}
Train (min) & 61.23  & 74.15  & 59.29  & 61.23   \\
Test (min) & 4.35  & 5.56  & 3.98  & 3.66 \\
\bottomrule
\label{tabel:dataset}
\end{tabular}
\end{table}
\begin{algorithm}[t]
\caption{Training strategy}
\label{alg:training}
\begin{algorithmic}[1]
\State $lr=2e-4$; $optimizer = Adam(lr, \ beta\_1=0.5)$;
\State $input \ length = 0.5s$; $\alpha = 1$; $\beta = 1$;
\State $epochs = 500$
\While{$input \ length <= max \ length$}
    \For{epoch in epochs}
        \If{epoch$>$(epochs $\times$ 65\%)} 
        \State $\alpha = 1$, $\beta = 0.5$;
        \State $lr += -5e-8$;
        \EndIf
    \EndFor
    \State $input \ length \ += 0.5s$;
     \State $input \ length \ = min(input \ length \ , max \ length)$;
    \State $\alpha = 1$, $\beta = 1$;
\EndWhile
\end{algorithmic}
\end{algorithm}

\begin{figure*}[t]
\centering
\includegraphics[width=.95\linewidth]{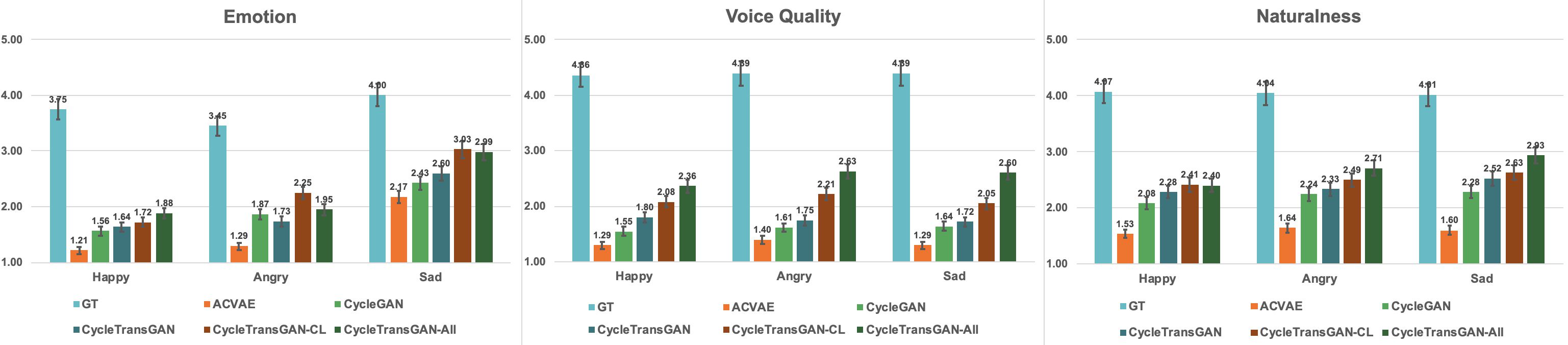}
\caption{Comparisons of MOS with 95\% confidence interval to evaluate emotion similarity, voice quality, and naturalness. GT indicates the ground truth.}
\label{fig:MOS}
\end{figure*}

The CycleGAN framework is incorporated with three types of loss: (1) consistency loss, (2) identity loss, and (3) adversarial loss. Thus, the loss functions for training the proposed model were defined as follows:
(1) Eq.\ref{cycleloss} demonstrates the cycle-consistency loss, where $x_a$ and $x_b$ are samples from A emotion and B emotion, respectively, $G_{A \rightarrow B}$ presents a generator to convert a sample from A to B, and $G_{B \rightarrow A}$ for B to A.  We calculated the L1 loss to compare the distance between the reconstructed sample and the original one, noted as $\Arrowvert \cdot \Arrowvert_1$. This loss is supposed to make the emotional information of the input consistent with the target one.
\begin{equation}
\begin{aligned}
\centering
L_{cyc}(G_{A \rightarrow B}, & G_{B \rightarrow A}) \\ 
& = E_{x_a}[\Arrowvert G_{B \rightarrow A}(G_{A \rightarrow B}(x_a))-x_a\Arrowvert_1] \\
&\ + E_{x_b}[\Arrowvert G_{A \rightarrow B}(G_{B \rightarrow A}(x_b))-x_b\Arrowvert_1]
\end{aligned}
\label{cycleloss}
\end{equation}
(2) Eq.\ref{idloss} introduces the identity loss, which encourages the generator to convert the input while retaining the original linguistic information.
\begin{equation}
\begin{aligned}
\centering
L_{id}(G_{A \rightarrow B}, G_{B \rightarrow A}) & = E_{x_a}[\Arrowvert G_{B \rightarrow A}(x_a)-x_a\Arrowvert_1] \\
&\ + E_{x_b}[\Arrowvert G_{A \rightarrow B}(x_b)-x_b\Arrowvert_1]
\end{aligned}
\label{idloss}
\end{equation}
(3) The adversarial loss is demonstrated as following equations, where $D_{A}$ and $D_{B}$ annotate discriminators for emotions A and B, respectively. The final adversarial loss is defined as $L_{adv} = L_{adv}^{A} + L_{adv}^{B}$. This loss attempts to tell whether a generator follows the target distribution.
\begin{equation}
\begin{aligned}
\centering
L_{adv}^{A}(G_{B \rightarrow A}, D_{A}) & = E_{x_a}[D_{A}(x_a)]\\
&\ + E_{x_b}[log(1-D_{A}({G_{A \rightarrow B}(x_b))}] \\
L_{adv}^{B}(G_{A \rightarrow B}, D_{B}) & = E_{x_b}[D_{B}(x_b)]\\
&\ + E_{x_a}[log(1-D_{B}({G_{B \rightarrow A}(x_a))}]
\end{aligned}
\label{advloss}
\end{equation}
Finally, the overall loss function is defined as:
\begin{equation}
\centering
L = L_{adv}+\alpha L_{cyc}+ \beta L_{id}
\label{floss}
\end{equation}
where $\alpha$ and $\beta$ are constants that will be defined during training. After converting spectrogram and F0, we employed WORLD vocoder \cite{morise2016world} to synthesize  the waveform.

\section{Experiment}

\subsection{Dataset}
% \begin{table}[t]
% \centering
% \caption{Information of dataset. The statistic data follow the format with \textit{duration/number of utterances}.}
% % For LaTeX tables use
% \begin{tabular}{lllll}
% \toprule
%  & Neutral & Happy &  Anger & Sad  \\
% \noalign{\smallskip}\hline\noalign{\smallskip}
% Train & 61.23min/1000 & 74.15min/1000 & 59.29min/1000 & 61.23min/1000  \\
% Test & 4.35min/70- & 5.56/70  & 3.98min/70 & 3.66min/70\\
% \bottomrule
% \label{tabel:dataset}
% \end{tabular}
% \end{table}

A Japanese emotional speech dataset \cite{asai2020emotional} that contains non-parallel happy, anger, sad, and neutral utterances was used in this study. 
Each category of this dataset has 1070 utterances in total. We assigned 1000 utterances to the training set and the left 70 utterances for the testing set, the duration of each emotion is presented in Table \ref{tabel:dataset}.
In this dataset, each sample has a similar corresponding one in other emotions. However, these similar samples are not paralleled. Samples of different emotions are used to express similar meanings with different expressions and modal particles.

\subsection{Settings and Baselines}

The hyper-parameters were set as Fig.\ref{fig:nn} shows\footnote{The implementation code and samples can be found at: \url{https://github.com/CZ26/CycleTransGAN-EVC}}. 
Moreover, we designed a training schedule with curriculum learning. As the Algorithm \ref{alg:training} demonstrates, we gradually increased the length of input with 0.5 second every 500 epochs to allow the model to see from the short speech to the long one. This strategy was supposed to introduce more detailed emotional features to the model.
Furthermore, the learning rate was decaying, and the weight constant $\beta$ was changed to 0.5 from 1 after 325 epochs, and reset back to 1 every 500 epochs.

As for baselines, we retrained the ACVAE \cite{kameoka2019acvae} and CycleGAN \cite{zhou2020transforming} on the Japanese emotional speech dataset and compare the performance with the proposed model.
Additionally, to verify the effects of the fine-grained level discriminator and curriculum learning, we modified our proposed to different configurations, i.e., CycleTransGAN (without curriculum learning and fine-grained level discriminator), CycleTransGAN-CL (with curriculum learning only), CycleTransGAN-All (with curriculum learning and fine-grained level discriminator).

\subsection{Results}

We adopted Mean Opinion Score (MOS) to subjectively evaluate emotional level, voice quality and naturalness of the samples. We invited 15 subjects to participate in all the experiments, and each subject evaluated 20 original utterances from the dataset and 75 converted utterances. 
Fig.~\ref{fig:MOS} presents the evaluation results with 95\% confidence interval. The ground truth is the score of original emotional samples.
For the scores of emotion evaluation, the CycleTransGAN-All model achieves the highest score for happy emotion, while the CycleTransGAN-CL achieves the highest score for angry and sad emotions.
As for the evaluation of voice quality and naturalness, the CycleTransGAN-all outperforms all the other models.
These results of CycleTransGAN-CL and CycleTransGAN imply that using curriculum learning improves emotional feature conversion by allowing the model to learn from the shorter segment up to the entire sample.
This is because the emotion of speech tends to appear in partly rather than in the whole speech, and the fine-grained level discriminator somewhat distracts the ability of the model to focus on the salient segment. Therefore, it leads to CycleTransGAN-All becomes inferior to CycleTransGAN-CL in terms of emotion similarity.

\section{Conclusion}
In this study, we proposed a CycleGAN-based emotional voice conversion model with a transformer, which is named CycleTransGAN. With the help of the transformer, the model is able to augment the receptive field to a wider range, which allows the generated speech to be more consistent in terms of temporal features, solving the instability problems of mispronunciations and skipped phonemes to some extent.
The experiment results show that the proposed models improve emotion similarity, voice quality, and naturalness. 
However, the clarity and emotional strength of the speech generated by the proposed models still need some further improvements. Future works will mainly focus on these parts.

\bibliographystyle{IEEEbib}
\small
\bibliography{refs}

\begin{thebibliography}{10}

\bibitem{helander2010voice}
Elina Helander, Tuomas Virtanen, Jani Nurminen, and Moncef Gabbouj,
\newblock ``Voice conversion using partial least squares regression,''
\newblock {\em IEEE Transactions on Audio, Speech, and Language Processing},
  vol. 18, no. 5, pp. 912--921, 2010.

\bibitem{aihara2012gmm}
Ryo Aihara, Ryoichi Takashima, Tetsuya Takiguchi, and Yasuo Ariki,
\newblock ``Gmm-based emotional voice conversion using spectrum and prosody
  features,''
\newblock {\em American Journal of Signal Processing}, vol. 2, no. 5, pp.
  134--138, 2012.

\bibitem{kawanami2003gmm}
Hiromichi Kawanami, Yohei Iwami, Tomoki Toda, Hiroshi Saruwatari, and Kiyohiro
  Shikano,
\newblock ``Gmm-based voice conversion applied to emotional speech synthesis,''
\newblock 2003.

\bibitem{ming2016exemplar}
Huaiping Ming, Dongyan Huang, Lei Xie, Shaofei Zhang, Minghui Dong, and Haizhou
  Li,
\newblock ``Exemplar-based sparse representation of timbre and prosody for
  voice conversion,''
\newblock in {\em 2016 IEEE International Conference on Acoustics, Speech and
  Signal Processing (ICASSP)}. IEEE, 2016, pp. 5175--5179.

\bibitem{takashima2013exemplar}
Ryoichi Takashima, Tetsuya Takiguchi, and Yasuo Ariki,
\newblock ``Exemplar-based voice conversion using sparse representation in
  noisy environments,''
\newblock {\em IEICE Transactions on Fundamentals of Electronics,
  Communications and Computer Sciences}, vol. 96, no. 10, pp. 1946--1953, 2013.

\bibitem{vekkot2020emotional}
Susmitha Vekkot, Deepa Gupta, Mohammed Zakariah, and Yousef~Ajami Alotaibi,
\newblock ``Emotional voice conversion using a hybrid framework with
  speaker-adaptive dnn and particle-swarm-optimized neural network,''
\newblock {\em IEEE Access}, vol. 8, pp. 74627--74647, 2020.

\bibitem{luo2016emotional}
Zhaojie Luo, Tetsuya Takiguchi, and Yasuo Ariki,
\newblock ``Emotional voice conversion using deep neural networks with mcc and
  f0 features,''
\newblock in {\em 2016 IEEE/ACIS 15th International Conference on Computer and
  Information Science (ICIS)}. IEEE, 2016, pp. 1--5.

\bibitem{robinson2019sequence}
Carl Robinson, Nicolas Obin, and Axel Roebel,
\newblock ``Sequence-to-sequence modelling of f0 for speech emotion
  conversion,''
\newblock in {\em ICASSP 2019-2019 IEEE International Conference on Acoustics,
  Speech and Signal Processing (ICASSP)}. IEEE, 2019, pp. 6830--6834.

\bibitem{kameoka2020convs2s}
Hirokazu Kameoka, Kou Tanaka, Damian Kwa{\'s}ny, Takuhiro Kaneko, and Nobukatsu
  Hojo,
\newblock ``Convs2s-vc: Fully convolutional sequence-to-sequence voice
  conversion,''
\newblock {\em IEEE/ACM Transactions on Audio, Speech, and Language
  Processing}, vol. 28, pp. 1849--1863, 2020.

\bibitem{choi2021sequence}
Heejin Choi and Minsoo Hahn,
\newblock ``Sequence-to-sequence emotional voice conversion with strength
  control,''
\newblock {\em IEEE Access}, vol. 9, pp. 42674--42687, 2021.

\bibitem{gao2018nonparallel}
Jian Gao, Deep Chakraborty, Hamidou Tembine, and Olaitan Olaleye,
\newblock ``Nonparallel emotional speech conversion,''
\newblock {\em arXiv preprint arXiv:1811.01174}, 2018.

\bibitem{ding2019group}
Shaojin Ding and Ricardo Gutierrez-Osuna,
\newblock ``Group latent embedding for vector quantized variational autoencoder
  in non-parallel voice conversion.,''
\newblock in {\em INTERSPEECH}, 2019, pp. 724--728.

\bibitem{zhou2020transforming}
Kun Zhou, Berrak Sisman, and Haizhou Li,
\newblock ``Transforming spectrum and prosody for emotional voice conversion
  with non-parallel training data,''
\newblock {\em arXiv preprint arXiv:2002.00198}, 2020.

\bibitem{liu2020emotional}
Songxiang Liu, Yuewen Cao, and Helen Meng,
\newblock ``Emotional voice conversion with cycle-consistent adversarial
  network,''
\newblock {\em arXiv preprint arXiv:2004.03781}, 2020.

\bibitem{cao2020nonparallel}
Yuexin Cao, Zhengchen Liu, Minchuan Chen, Jun Ma, Shaojun Wang, and Jing Xiao,
\newblock ``Nonparallel emotional speech conversion using vae-gan.,''
\newblock in {\em INTERSPEECH}, 2020, pp. 3406--3410.

\bibitem{moritani2021stargan}
Asuka Moritani, Ryo Ozaki, Shoki Sakamoto, Hirokazu Kameoka, and Tadahiro
  Taniguchi,
\newblock ``Stargan-based emotional voice conversion for japanese phrases,''
\newblock {\em arXiv preprint arXiv:2104.01807}, 2021.

\bibitem{dosovitskiy2020image}
Alexey Dosovitskiy, Lucas Beyer, Alexander Kolesnikov, Dirk Weissenborn,
  Xiaohua Zhai, Thomas Unterthiner, Mostafa Dehghani, Matthias Minderer, Georg
  Heigold, Sylvain Gelly, et~al.,
\newblock ``An image is worth 16x16 words: Transformers for image recognition
  at scale,''
\newblock {\em arXiv preprint arXiv:2010.11929}, 2020.

\bibitem{wu2020transformer}
Chuhan Wu, Fangzhao Wu, and Yongfeng Huang,
\newblock ``Da-transformer: Distance-aware transformer,''
\newblock {\em arXiv preprint arXiv:2010.06925}, 2020.

\bibitem{kameoka2019acvae}
Hirokazu Kameoka, Takuhiro Kaneko, Kou Tanaka, and Nobukatsu Hojo,
\newblock ``Acvae-vc: Non-parallel voice conversion with auxiliary classifier
  variational autoencoder,''
\newblock {\em IEEE/ACM Transactions on Audio, Speech, and Language
  Processing}, vol. 27, no. 9, pp. 1432--1443, 2019.

\bibitem{ming2015fundamental}
Huaiping Ming, Dongyan Huang, Minghui Dong, Haizhou Li, Lei Xie, and Shaofei
  Zhang,
\newblock ``Fundamental frequency modeling using wavelets for emotional voice
  conversion,''
\newblock in {\em 2015 International Conference on Affective Computing and
  Intelligent Interaction (ACII)}. IEEE, 2015, pp. 804--809.

\bibitem{kaneko2018cyclegan}
Takuhiro Kaneko and Hirokazu Kameoka,
\newblock ``Cyclegan-vc: Non-parallel voice conversion using cycle-consistent
  adversarial networks,''
\newblock in {\em 2018 26th European Signal Processing Conference (EUSIPCO)}.
  IEEE, 2018, pp. 2100--2104.

\bibitem{morise2015cheaptrick}
Masanori Morise,
\newblock ``Cheaptrick, a spectral envelope estimator for high-quality speech
  synthesis,''
\newblock {\em Speech Communication}, vol. 67, pp. 1--7, 2015.

\bibitem{morise2016world}
Masanori Morise, Fumiya Yokomori, and Kenji Ozawa,
\newblock ``World: a vocoder-based high-quality speech synthesis system for
  real-time applications,''
\newblock {\em IEICE TRANSACTIONS on Information and Systems}, vol. 99, no. 7,
  pp. 1877--1884, 2016.

\bibitem{asai2020emotional}
Sara Asai, Koichiro Yoshino, Seitaro Shinagawa, Sakriani Sakti, and Satoshi
  Nakamura,
\newblock ``Emotional speech corpus for persuasive dialogue system,''
\newblock in {\em Proceedings of The 12th Language Resources and Evaluation
  Conference}, 2020, pp. 491--497.

\end{thebibliography}

\end{document}